\documentclass[aps,prb,twocolumn,showpacs,superscriptaddress,floatfix]{revtex4}
\usepackage{graphics,amssymb,amsmath,epsfig,psfrag}
\newcommand{\be}[0]{\begin{eqnarray}}
\newcommand{\ee}[0]{\end{eqnarray}}
\newcommand{\nn}{\nonumber}

\begin{document}

\title{Defect Motion and Lattice Pinning Barrier in Josephson-Junction Ladders}

\author{H. Kang}
\affiliation{Department of Physics and Center for Theoretical Physics,
Seoul National University, Seoul 151-747, Korea}

\author{Jong Soo Lim}
\affiliation{Department of Physics and Center for Theoretical Physics,
Seoul National University, Seoul 151-747, Korea}

\author{J.-Y. Fortin}
\affiliation{Laboratoire de Physique Th\'eorique, Universit\'e Louis Pasteur,
67084 Strasbourg, France}

\author{J. Choi}
\affiliation{Department of Physics, Keimyung University,
Daegu 704-701, Korea}

\author{M.Y. Choi}
\affiliation{Department of Physics and Center for Theoretical Physics,
Seoul National University, Seoul 151-747, Korea}
\affiliation{Korea Institute for Advanced Study, Seoul 130-722, Korea}

\begin{abstract}
We study motion of domain wall defects in a fully frustrated Josephson-junction ladder system,
driven by small applied currents.
For small system sizes, the energy barrier $E_B$ to the defect motion is computed
analytically via symmetry and topological considerations.
More generally, we perform numerical simulations directly on the equations of motion,
based on the resistively-shunted junction model,
to study the dynamics of defects, varying the system size.
Coherent motion of domain walls is observed for large system sizes.
In the thermodynamical limit, we find $E_B=0.1827$ in units of the Josephson coupling energy.
\end{abstract}


\maketitle

\section{Introduction}

Two-dimensional (2D) arrays of Josephson junctions are of interest in various fields of
fundamental classical and quantum physics.
In the simplest case, they provide an experimental realization of the $XY$ model;
in particular, applying a magnetic field introduces frustration,
measured by the flux per plaquette in units of the flux quantum.\cite{FXY}
The corresponding vortices induced by the field tend to form a regular
flux lattice, thus lowering the free energy,
and result in interplay with the underlying lattice periodicity.
This gives rise to commensurate-incommensurate effects and leads
to a rich variety of physics, including first-order and double
transitions, reentrance, glassy behavior, quantum transitions,
topological quantization, dynamic transitions and resonance,
etc.\cite{review,dyn}
In these phenomena vortex configurations and dynamics play crucial roles,
driving transitions and governing transport properties.
Here one interesting question arises when an extra vortex is added into the system.
While the vortex in general sits on a plaquette with minimum energy,
which is separated by the potential barrier set by the underlying lattice structure,
it may be driven into motion by applying currents,
as it is exerted by the ``Lorentz force'' in the transverse direction,
and accordingly generates non-vanishing voltage.
Indeed, the voltage measurement in recent dynamic simulations,\cite{defect}
performed in the presence of external currents,
has given the pinning energy barriers as well as the critical currents,
which agree fully with experimental results,\cite{exp} thus resolved the long-standing
discrepancy in the frustrated case.

This paper focuses on the vortex dynamics in ladders of Josephson junctions,
which provides the simplest system for probing the frustration effects:
Those studied in existing literature include the vortex configuration and the critical current,
depending on the frustration,\cite{Mazo,Stroud}
the vortex-vortex interaction decaying exponentially,\cite{Kim}
quantum effects\cite{quantum}, and resonance.\cite{SR}
Note the vast difference from the 2D system, especially, in the vortex interaction,
which is expected to affect significantly the dynamics of a vortex in a background of
other vortices, i.e., in a frustrated system.
In particular domain walls in a ladder system assume the simple form of point defects,
the dynamics of which is convenient to probe.
We thus consider the domain wall defects created by
adding an extra vortex in a fully frustrated ladder and examine their motion
driven by external currents.
In small systems, the symmetry argument and topological constraints allow one
to computed analytically the energy barrier.
More generally, the defect motion, driven by uniform currents, is investigated
by means of dynamical simulations performed directly on the equations of motion.
The resulting value of the energy barrier is found consistent with the analytical one
obtained for small systems.
Also observed is the defect motion, either sequential or simultaneous,
depending on the size and the initial configuration.
Such characteristics are attributed to the distance-dependent interaction between
defects and the underlying lattice geometry.

There are five sections in this paper: Section II introduces the model system whereas
Sec. III is devoted to the analytical calculations of the energy barrier to the
defect motion in small systems.  In Sec. IV, we describe the numerical simulations
performed on the equations of motion in the presence of uniform driving currents,
and present the results. The current-voltage ($IV$) characteristics and the energy function
are computed, which in turn give the critical current and the pinning energy barrier
for various system sizes.  Finally, a brief summary is given in Sec. V.

\section{Model System}

\begin{figure}
\epsfig{width=0.45\textwidth,file=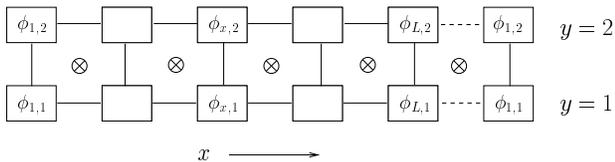}
\caption{Schematic notation for a Josephson-junction ladder.
Each superconducting grain, denoted by a square, is characterized by
the phase $\phi_{x,y}$ of the superconducting order parameter.
The symbol $\otimes$ denotes the flux per plaquette from
an external transverse magnetic field.
The extra plaquette on the right hand side represents the periodic
boundary conditions.}
\label{ladder}
\end{figure}

We consider a ladder of Josephson junctions made of $2L$ superconducting grains
weakly coupled to their nearest neighbors, the schematic diagram of which is shown
in Fig.~\ref{ladder}.
The grains are located at sites $i \equiv (x, y)$,
where $x$ runs from $1$ to $L$ (in the leg direction) and the label $y\,(=1, 2)$
describes respectively the lower and upper legs of the ladder.
Each grain is characterized by the local condensed wave function or the order parameter:
\begin{equation}
\Psi_{i} = |\Psi_{i}| e^{i\phi_{i}}.
\end{equation}
where the local superconducting fluid density $|\Psi_{i}|$ is assumed to be
constant at low temperatures.
Accordingly, relevant fluctuations come from the phases $\phi_{i}$
and the Hamiltonian of the system in the presence of the
external field is simply given by the sum of the nearest neighboring pair
energies
\begin{equation}
H = -E_J \sum_{\langle i, j\rangle} \cos(\phi_i -\phi_j -A_{ij}),
\end{equation}
where $E_J$ is the coupling constant between the grains,
$\langle i, j \rangle$ represents nearest neighboring pairs,
and the bond angle $A_{ij}$ is given by the line integral of the vector potential:
\begin{equation}
A_{ij} = \frac{2\pi}{\Phi_0} \int_i^j {\bf A}\cdot d{\bf l}
\end{equation}
with the flux quantum $\Phi_0 \equiv \pi\hbar c/e$.
In the Landau gauge, the components of the vector potential ${\bf A}(x, y)$ are given by
\be
A_x(x,y)=0 ~~ \mbox{and} ~~A_y(x,y)= \Phi x ,
\ee
where $\Phi$ is the magnetic flux per plaquette
and $x$ is the position along the leg direction.

For the ladder in Fig.~\ref{ladder}, the Hamiltonian reduces to
\be
H = &-& E_J \sum_{x, y} \cos(\phi_{x,y} - \phi_{x{+}1,y}) \nn \\
 &-& E_J \sum_{x} \cos(\phi_{x,1} - \phi_{x,2} - 2\pi fx),
\label{energy}
\ee
where $f \equiv \Phi/\Phi_0$ measures the frustration of the system.
In the fully frustrated case ($f=1/2$), which is our main concern in this work,
one every other site is occupied by a single vortex.

\begin{figure}
\epsfig{width=0.45\textwidth,file=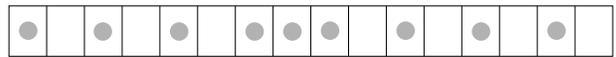}
\caption{Vortex configuration in the presence of an extra vortex in the fully
frustrated ladder of size $L=16$.
Filled circles represent vortices.}
\label{defect}
\end{figure}

We now add or remove one vortex; this creates topological defects (domain walls) that
affect the ground state.
A typical vortex configuration in this case is displayed in Fig.~\ref{defect}.
The extra vortex can move through the periodic potential
produced by the lattice structure when it is subject to a perpendicular current.
An estimation of the corresponding lattice pinning barrier is then
made each time this extra vortex crosses the barrier.
Note that the periodic potential is in general modulated significantly by other (underlying) vortices
present in the system with $f=1/2$, resulting in the barrier strikingly different from that
in the unfrustrated system ($f=0$).

\section{Analytic Calculations}

For convenience, we choose new gauge invariant phases that simplify
the Hamiltonian and the current distribution in the system.
Let $\theta_x$ and $\varphi_x$ denote the following phase differences
between the grains:
\be 
\theta_x &=& \phi_{x+1,1} - \phi_{x,1} \nn \\
\theta'_x &=& \phi_{x+1,2} - \phi_{x,2} \nn \\
\varphi_x &=& \phi_{x,1} - \phi_{x,2} - 2\pi f x.
\ee
It is easy to see, by symmetry and energy considerations, that the phase differences
$\theta_x$ and $\theta'_x$ are opposite to each other.\cite{Mazo}
Indeed, the sum of these phases around each plaquette is constrained topologically by the
flux or frustration $f$ and the (integer) vortex number $n_x$: 
\be
\theta_x-\theta'_x-\varphi_{x+1}+\varphi_{x}=2\pi(n_x-f)\equiv 2\pi q_x ,
\ee
where $q_x$ is the (fractional) vortex charge,
and the Hamiltonian simply reads
\be \nn
H&=&-E_J\sum_{x}\left[\cos\theta_x +\cos\theta'_x +\cos\varphi_x \right] \\
&=&-E_J\sum_x \left\{2\cos\left[\frac{\varphi_{x+1}-\varphi_{x}}{2} + \pi(n_x -f)\right] \right.
\nn \\
& &\left. ~~~~~~~~~~~~\times \cos\left(\frac{\theta_x+\theta'_x}{2}\right) +\cos\varphi_x \right\}.
\ee
Then the condition $\theta'_x=-\theta_x$ decouples the phases between the transverse directions
and leads to a solution that minimizes this Hamiltonian.
%
Using the current conservation laws, we can write a set of $L$
equations for $\theta_x$ and $\varphi_x$ at every node of the lattice: 
\be\label{eqph} 
\sin\theta_x =\sin\theta_{x+1}-\sin\varphi_{x+1} 
\ee 
with the boundary conditions 
\be 
\theta_{x+L} = \theta_x ~~\mbox{and}~~ \varphi_{x+L}=\varphi_x . 
\ee
%

\begin{figure}
\psfrag{T1}{$\theta$}
\psfrag{T2}{$\theta '$}
\psfrag{A}{$\varphi$}
\psfrag{AA}{$(a)$}
\psfrag{AAA}{$(b)$}
\includegraphics[scale=0.3]{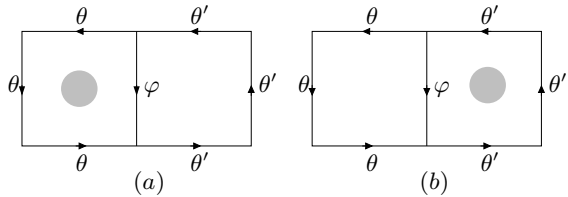}
\caption{Configuration of a two-plaquette system, with the phase difference labeled along each link.
Filled circles represent vortices.}
\label{example}
\end{figure}

The barrier energy $E_B$ for a vortex moving along a ladder can be computed
exactly on one simple example. In Fig.~\ref{example},
we consider two plaquettes under closed boundary conditions and a single vortex in the system.
In the notation of Fig.~\ref{example}, the equations for the phases
$(\theta,\theta',\varphi)$ in case (a) are given by
\be \nn
3\theta-\varphi &=&\pi \\
3\theta'+\varphi &=&-\pi \nn \\
\sin\theta' &=&\sin\theta +\sin\varphi ,
\ee
which yields $\theta=-\theta'=(\pi+\varphi)/3$.
As a function of $\varphi$, the energy
\be
E(\varphi)=-6\cos\left(\frac{\pi+\varphi}{3}\right) -\cos\varphi
\ee
has an absolute minimum for $\varphi=-\pi/2$, which in turn leads to $\theta=\pi/6$ and
$E=-3\sqrt{3}\approx -5.196$, and a maximum for $\varphi=-\pi$
together with $\theta=0$ and $E=-5$.
On the other hand, in case (b), we have
\be \nn
3\theta-\varphi &=&-\pi \\
3\theta'+\varphi &=&\pi \nn \\
\sin\theta' &=&\sin\theta +\sin\varphi , 
\ee 
the solutions of which are $\varphi=\pi/2$ and $\theta=-\pi/6$ for the ground state
(with $E=-3\sqrt{3}$) and $\varphi=\pi$ and $\theta=0$ for the
excited state (with $E=-5$). The excited states in both cases are
equivalent since $\varphi=-\pi=\pi\,(\mbox{mod}\,2\pi)$. This
corresponds to the situation that the system evolves from
configuration (a) to (b), namely, the instant when the vortex is
exactly on the rung between the two plaquettes. 
Accordingly, the energy barrier is simply given by 
\be 
E_B=3\sqrt{3}-5 \approx 0.196.
\ee

In the general case, the value $\varphi=\pi$ (or $-\pi$) in the excited state does not depend on
the frustration parameter $f$ since it is always a solution of the equation
$\partial E(\varphi)/\partial\varphi =0$ with
\be \nn
E(\varphi)=&-&3\cos\left(\frac{2\pi f+\varphi}{3}\right)
-3\cos\left[\frac{2\pi (1-f)+\varphi}{3}\right] \\
&-&\cos\varphi .
\ee
In the following, we accept that $\varphi=\pi$ corresponds to the
solution of the excited state in which the vortex is on the rung for any given $L$;
this will be checked numerically (see Fig.~\ref{phi6} below). 

\begin{figure}
\epsfig{width=0.45\textwidth,file=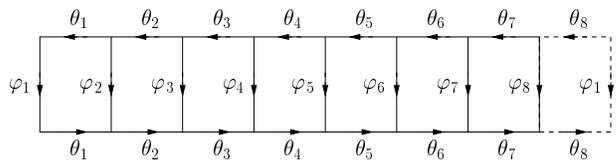}
\caption{Phase configuration of a Josephson-junction ladder of
$L=8$ plaquettes. Periodic boundary conditions are employed.}
\label{phc}
\end{figure}

\begin{figure}
\epsfig{width=0.45\textwidth,file=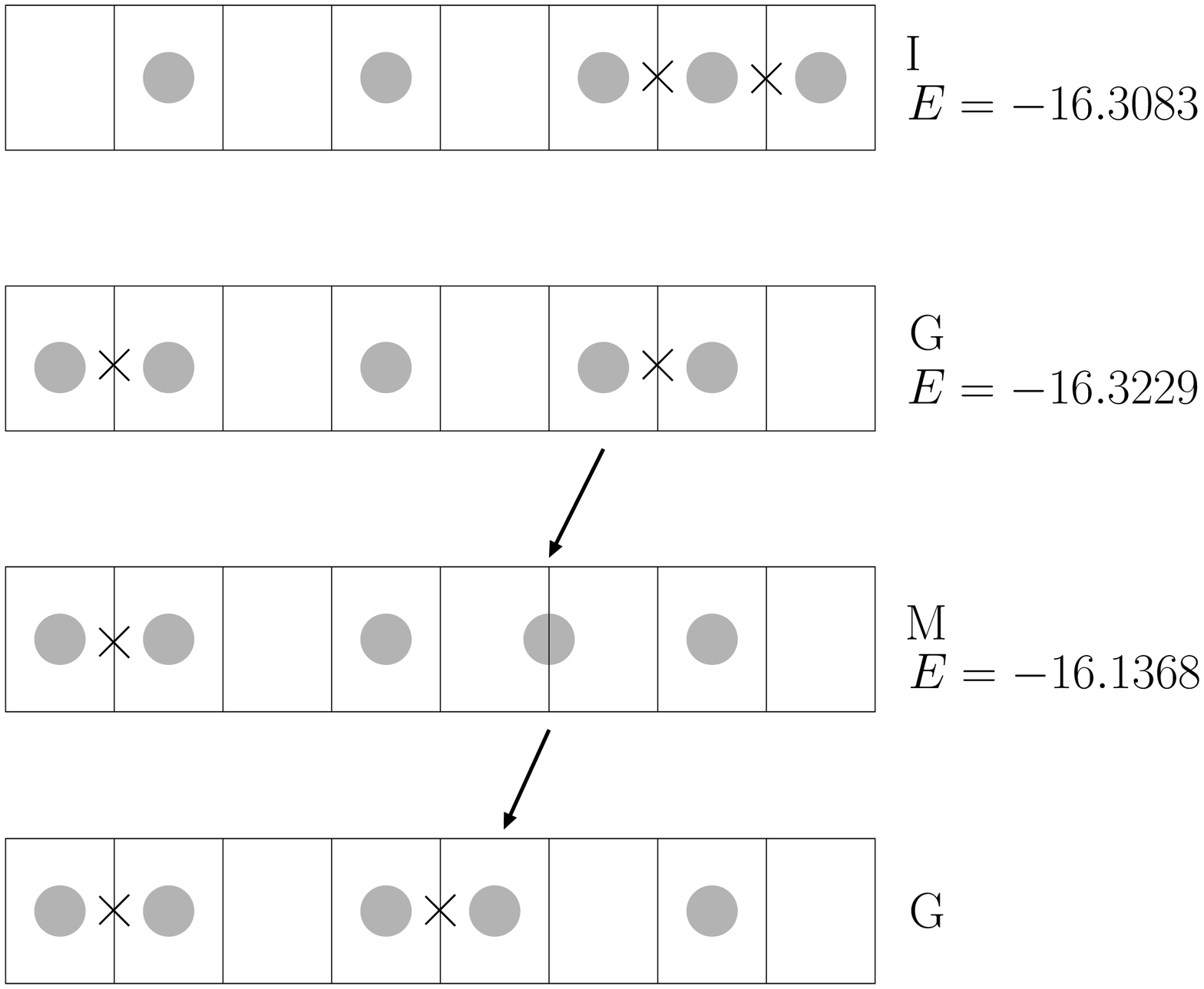}
\caption{Three vortex configurations (I,\,G,\,M), showing the presence of vortices
(denoted by filled circles) and domain wall defects (denoted by crosses).
Also shown is the estimated energy (in units of $E_J$) of each configuration.
Arrows represent the time evolution of the configuration, which
has been verified by extensive numerical simulations.}
\label{thconf}
\end{figure}

In a more complicate case, we consider a system with $L=8$ plaquettes.
Figure~\ref{phc} shows the phase configuration of an eight-plaquette
system at $f=1/2$ under periodic boundary conditions, in the presence of an extra vortex.
We consider three possible configurations (I,\,G,\,M) shown in Fig.~\ref{thconf},
where filled circles and crosses represent vortices and defects (domain walls), respectively.
Starting from the initial state I and driven by the injected current along the $y$ direction,
the system evolves eventually to configuration G via a number of intermediate configurations.
It subsequently evolves to M and back to G.
Configurations G and M correspond to the lowest-energy state and the high-energy (excited) state,
respectively, and this evolution pattern repeats with time, which
has been verified by extensive numerical simulations.

In the initial configuration I, the stationary phase relations are given by
\begin{align}
2\theta_1 - \varphi_2 +\varphi_1 &= -\pi,
& 2\theta_2 - \varphi_3 + \varphi_2 &= \pi \nonumber \\
2\theta_3 - \varphi_4 +\varphi_3 &= -\pi,
& 2\theta_4 - \varphi_5 + \varphi_4 &= \pi \nonumber \\
2\theta_5 - \varphi_6 +\varphi_5 &= -\pi,
& 2\theta_6 - \varphi_7 + \varphi_6 &= \pi \nonumber \\
2\theta_7 - \varphi_8 +\varphi_7 &= \pi,
& 2\theta_8 - \varphi_1 + \varphi_8 &= \pi.
\end{align}
In units of the Josephson coupling energy $E_J$, the energy
is estimated to be $E(I) = -16.3083$.
For configuration G, the phase relations read
\begin{align}
2\theta_1 - \phi_2 +\varphi_1 &= \pi,
& 2\theta_2 - \varphi_3 + \varphi_2 &= \pi \nonumber \\
2\theta_3 - \phi_4 +\varphi_3 &= -\pi,
& 2\theta_4 - \varphi_5 + \varphi_4 &= \pi \nonumber \\
2\theta_5 - \phi_6 +\varphi_5 &= -\pi,
& 2\theta_6 - \varphi_7 + \varphi_6 &= \pi \nonumber \\
2\theta_7 - \phi_8 +\varphi_7 &= \pi,
& 2\theta_8 - \varphi_1 + \varphi_8 &= -\pi,
\end{align}
which yields the energy $E(G)= -16.3229$.

Configuration M describes an intermediate state via which
the system goes from the state with the occupation number ($n_5=0,\,n_6=1$) to
that with ($n_5= 1,\,n_6=0$),
namely, the vortex moves to the left by one plaquette,
similarly to the evolution from (b) to (a) in Fig.~\ref{example}.
In this case, the vortex numbers in both cells are not well defined,
but the vortex is said to be ``spread'' between the two plaquettes.
To apply Eq.~\eqref{eqph}, we further take the two plaquettes on both sides
of the rung as one unit cell. Since the net vortex charge enclosed in this cell
(consisting of the two plaquettes) is zero, the sum of phase differences around it
also vanishes. We thus have the condition
$2\theta_5+\varphi_5+2\theta_6-\varphi_7=2\pi(n_5-f+n_6-f)=0$, with
$n_5+n_6=1$.
The remaining relations are given by
\begin{align}
2\theta_1 - \varphi_2 +\varphi_1 &= \pi,
& 2\theta_2 - \varphi_3 + \varphi_2 &= \pi \nonumber \\
2\theta_3 - \varphi_4 +\varphi_3 &= -\pi,
& 2\theta_4 - \varphi_5 + \varphi_4 &= \pi, \nonumber \\
%
2\theta_7 - \varphi_8 +\varphi_7 &= \pi
& 2\theta_8 - \varphi_1 + \varphi_8 &= -\pi.
\end{align}
As addressed already, the vortex sits on the rung in this configuration and the phases
take the radial direction around the center of the rung, thus leading to the phase difference
$\varphi_6 =\pi$ along the rung. 
This is manifested by the time evolution of $\varphi_6$, as shown in the next section (see Fig.~\ref{phi6}).
We thus set $\varphi_6=\pi$ and obtain the energy of the configuration: $E(M)=-16.1368$.

Together with the result of $E(G)$, we estimate the pinning barrier according to
\begin{equation}
E_B\equiv E(M)-E(G)=0.1861.
\end{equation}
Note that this value, obtained for $L=8$, is lower than the value $0.19615$
in the two-plaquette case ($L=2$).
We thus expect that the energy barrier $E_B$ in the thermodynamic limit
($L\rightarrow \infty$) has a value still lower than $0.1861$.

\section{Numerical Simulations}

To evaluate the precise value of the energy barrier for various system sizes,
we have performed extensive dynamic simulations on the resistively shunted junction (RSJ) model.
The dynamics of the RSJ model, with single-junction critical current $i_c$ and shunt
resistance $R$, is governed by the set of equations of motion for the phase $\phi_i$,
\begin{equation}
{\sum_j}' \left[\frac{\hbar}{2eR}\frac{ d\widetilde{\phi}_{ij}} {dt} +
i_c \sin\widetilde{\phi}_{ij}\right] = I_i ,
\label{rsj}
\end{equation}
where $\widetilde{\phi}_{ij}\equiv \phi_i-\phi_j -A_{ij}$ is the gauge-invariant phase difference across
the junction $(ij)$, and the primed summation runs over the nearest neighbors
of grain $i$.
The system is driven by the current $I_i = I_{x,y} =I (\delta_{y,2} -\delta_{y,1})$ (applied to
grain $i$), namely, uniform current $I$ is injected to and extracted from each grain on the
upper ($y=2$) and lower ($y=1$) legs, respectively.
Using a modified Euler method,
we have integrated Eq.~\eqref{rsj} with the time step of size
$\Delta t = 0.05$ (in units of $\hbar/2ei_cR$)
for a variety of ladders up to the system size $L=512$.
In addition to the periodic boundary conditions imposed along the $x$ direction,
we introduce a $2\pi$ phase slip across the whole system:
\begin{eqnarray}
\phi_{L+1,2} &=& \phi_{1,2} + \pi \nn\\
\phi_{L+1,1} &=& \phi_{1,1} - \pi ,
\end{eqnarray}
which generates a single extra vortex.

\begin{figure}
\epsfig{width=0.42\textwidth,file=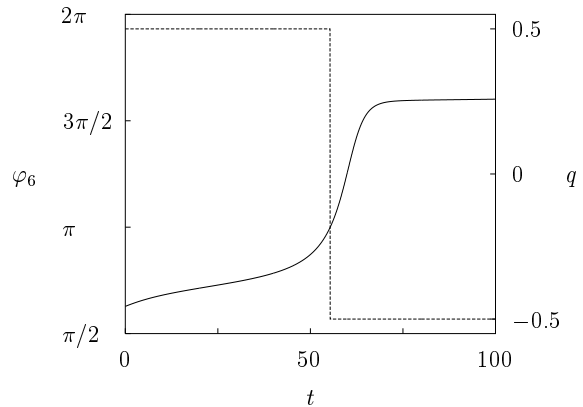}
\caption{Time evolution of the phase difference $\varphi_6$ across the rung (solid line,
left vertical axis), together with that of the vortex charge $q = n-f$ on the plaquette
just right of the rung (dotted line, right vertical axis).
Time $t$ is given in units of $\hbar/2ei_c R$.
}
\label{phi6}
\end{figure}

We first examine how the rung phase difference $\varphi_6$ varies in the vortex motion
and plot in Fig.~\ref{phi6} its time evolution in the system of eight plaquettes.
Also plotted is the evolution of the vortex charge $q = n-f$ (with $n$ being the vortex number)
on the plaquette just right of the rung.
It is observed that $q$ (or $n$) changes rather abruptly from $1/2$ to $-1/2$
(or from $1$ to $0$), describing the motion of a vortex to the left.
In particular, at the moment of the change, i.e., when the vortex is located {\em on} the rung,
the phase difference $\varphi_6$ indeed has the value $\pi$, as expected.

\begin{figure}
\epsfig{width=0.4\textwidth,file=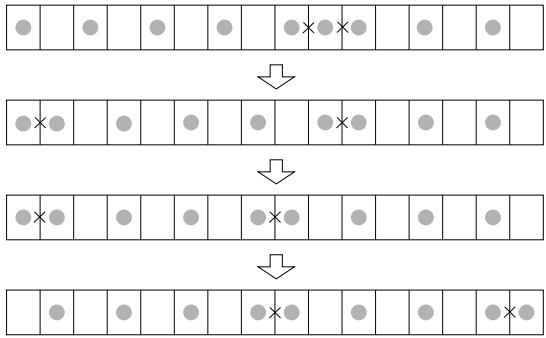}
\caption{Pattern of defect motion in a fully frustrated ladder ($f=1/2$),
with filled circles and crosses denoting vortices and domain walls, respectively,
as time goes by (in the direction of the arrows).
Currents are applied uniformly along the rungs.
}
\label{evolution}
\end{figure}

\begin{figure}
\epsfig{width=0.4\textwidth,file=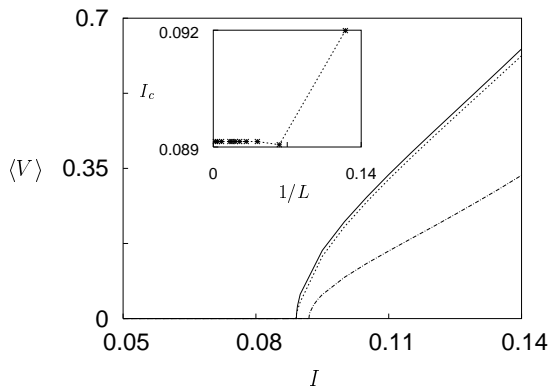}
\caption{$IV$ characteristics for the system size $L=8$, $16$, and $24$, respectively,
from bottom to top.  The inset shows the critical current $I_c$ as a function of $L$.
Current $I$ and voltage $V$ are expressed in units of $i_c$ and $i_c R$, respectively.}
\label{iv}
\end{figure}

Figure~\ref{evolution} shows typical motion of defects under the driving currents.
At first, two defects (i.e., two domain walls separating the three neighboring vortices)
are next to each other, as shown in the first configuration (from top to bottom).
The distance between the domain wall defects grows with time until this distance eventually
becomes half the system size (see the second configuration).
Then, the defect on the right moves first (changing the configuration to the third one),
subsequently followed by the motion of the one on the left (resulting in the fourth configuration).
In the case that there are only a few plaquettes ($L<40$), this behavior
is always observed, regardless of the initial distance between the two defects.
On the other hand, in a system of larger size, two types of
behavior are observed, depending on initial conditions:
When the two defects are initially located at nearby sites,
they move simultaneously through transient states and the distance between them
does not grow beyond $20$ plaquettes.
In contrast, two defects distant by more than $20$ plaquettes
tend to move sequentially for appropriate initial phase configurations.
We presume that such size dependence has its origin in the interaction
between defects and the underlying periodic lattice geometry.
Namely, the interaction between two domain wall defects becomes vanishingly small
as the distance is increased beyond $20$ plaquettes,
which may reflect the exponentially decaying interaction between vortices.\cite{Kim}
In this manner the characteristic interaction between domain walls in a background of vortices
appears to be exposed.

\begin{figure}[b]
\epsfig{width=0.4\textwidth,file=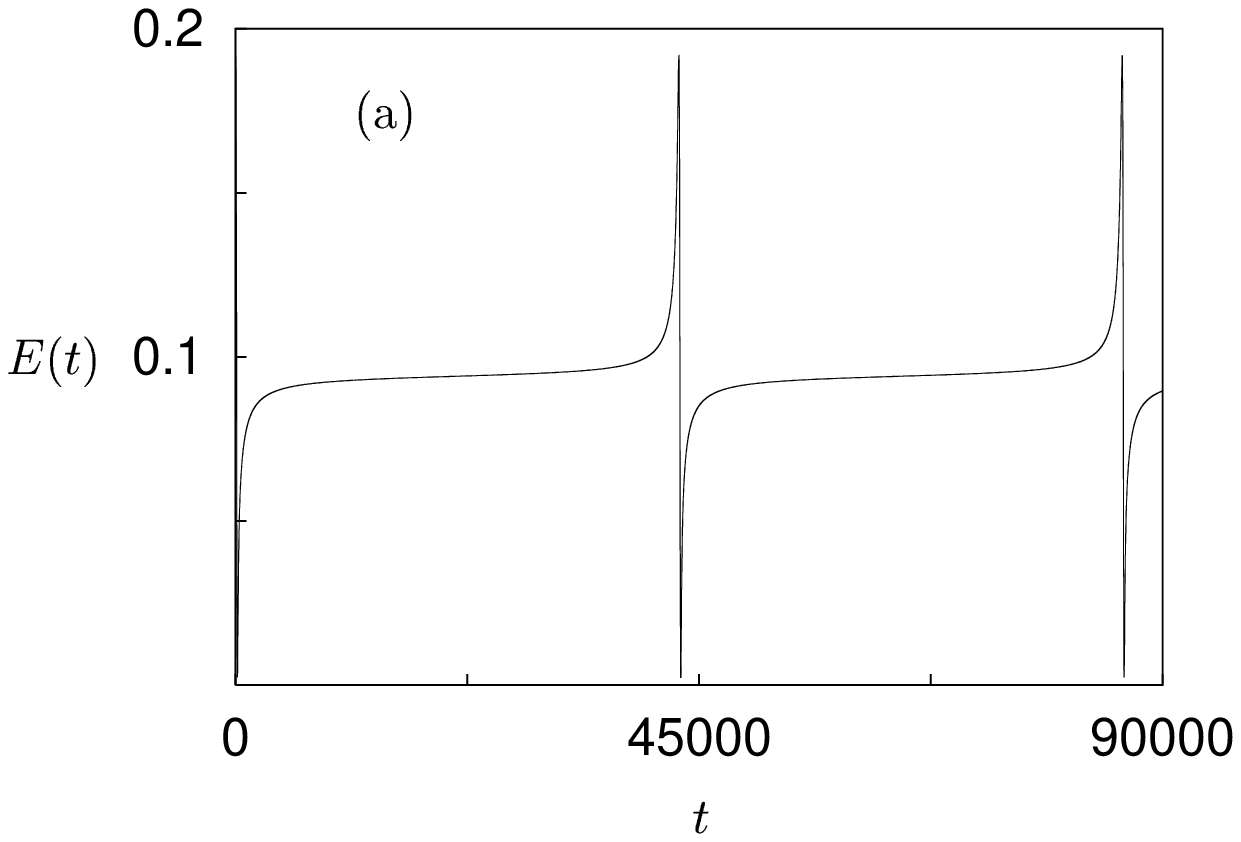}
\epsfig{width=0.4\textwidth,file=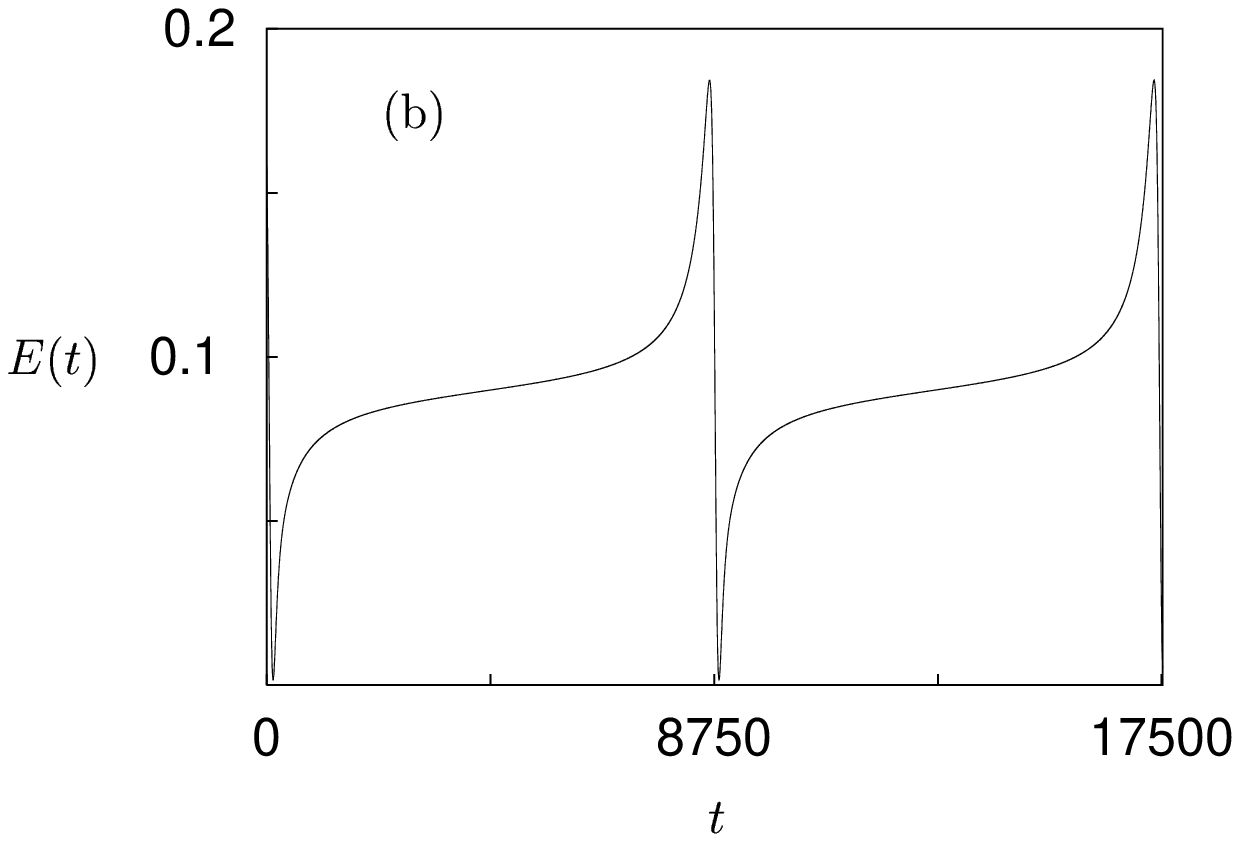}
\epsfig{width=0.4\textwidth,file=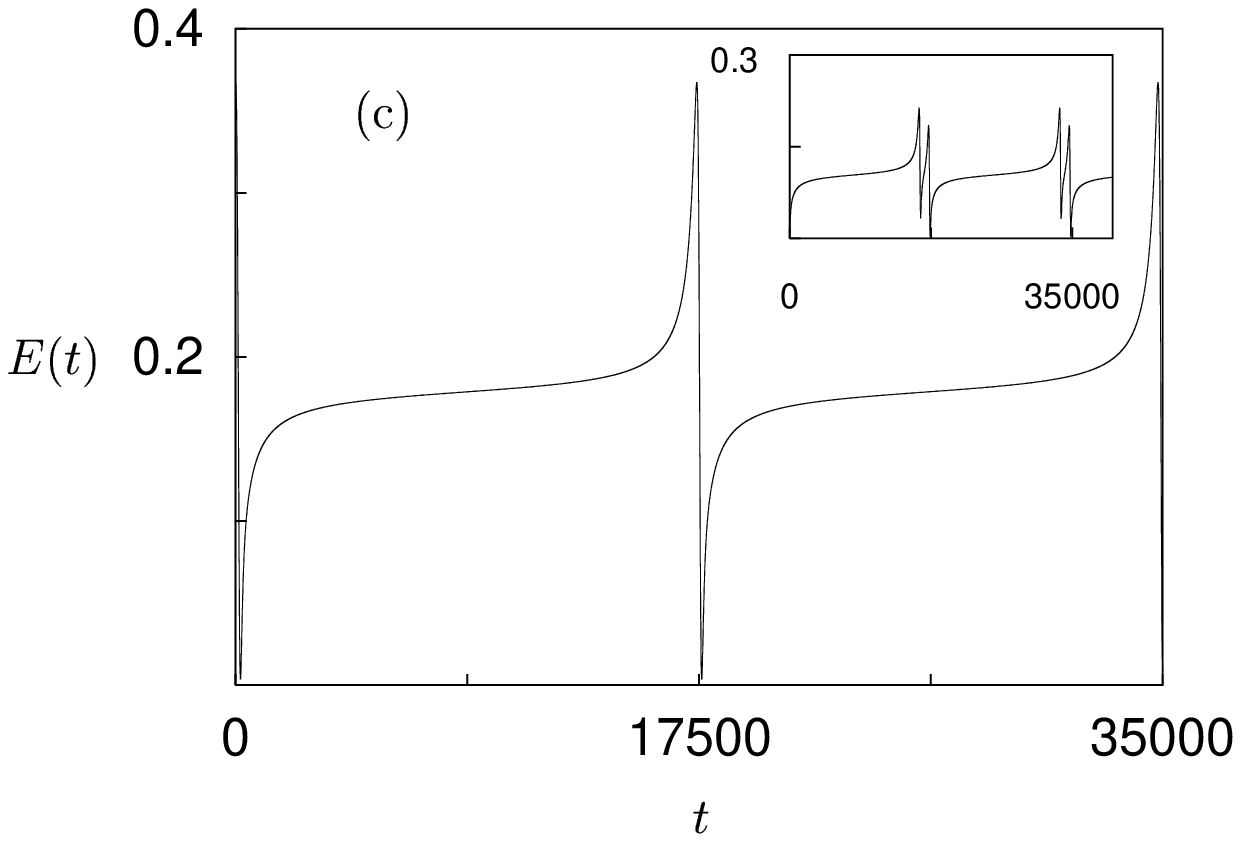}
\caption{Time evolution of energy $E(t)$ in systems of size $L= 8$ (a) and 64
[(b) and (c)]. The uniform driving current
$I=I_c(L)+0.0001$ has been applied along each rung.
(a) and (b) describe the {\it sequential} motion while
(c) corresponds to the {\it simultaneous} motion (see the text).
For convenience, $E(t)$, given in units of $E_J$, has been shifted 
such that $E=0$ corresponds to the minimum.
The inset in (c) shows a transient behavior:
The two peaks merge eventually into one peak shown in the main plate.}
\label{evt}
\end{figure}

\begin{figure}
\epsfig{width=0.4\textwidth,file=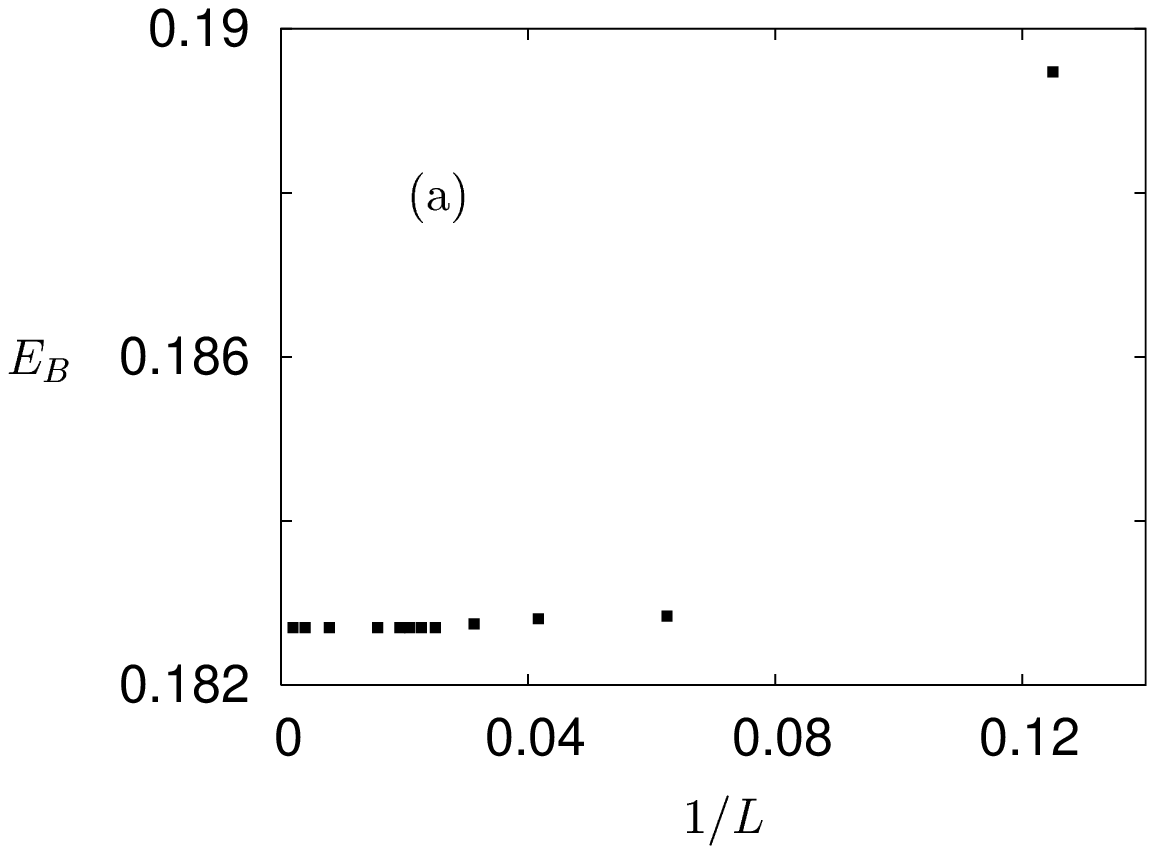}
\epsfig{width=0.4\textwidth,file=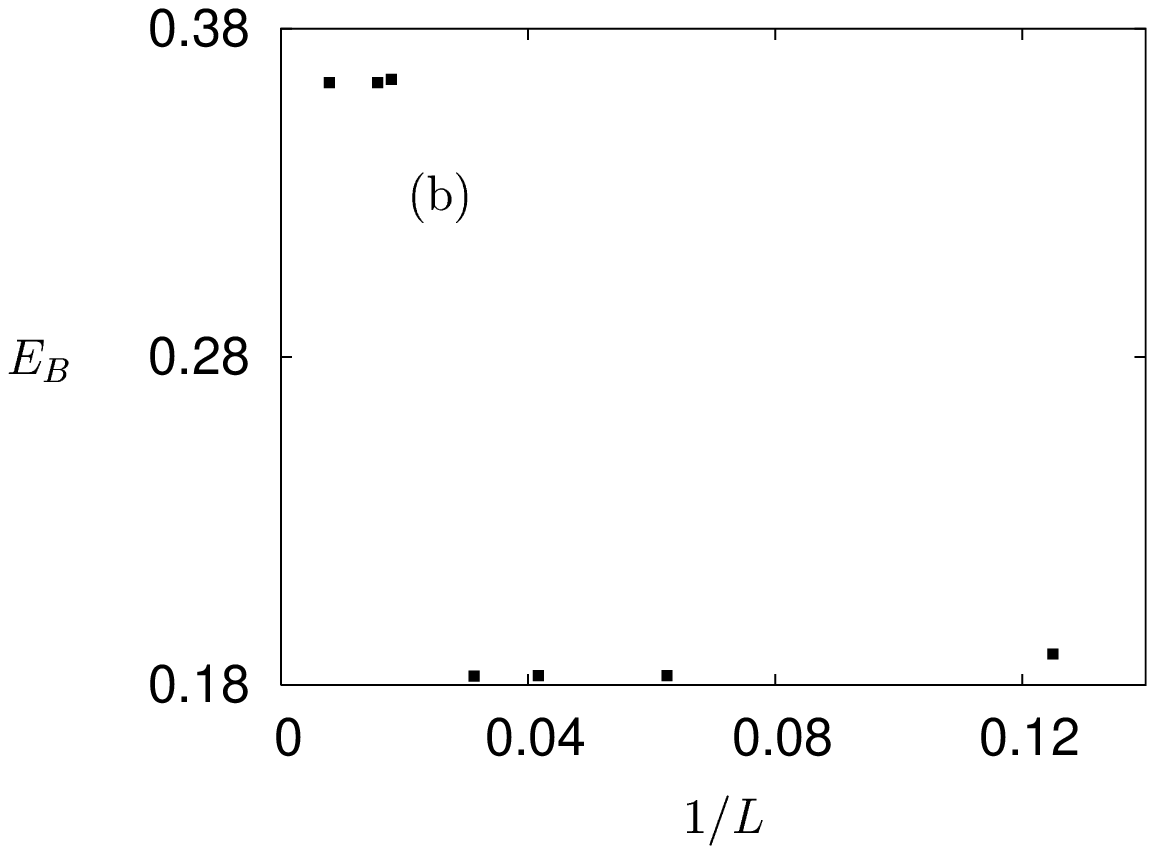}
\caption{Pinning energy barrier $E_B$ (in units of $E_J$) as a function of the system size $L$.
Energy barriers corresponding to (a) sequential and (b)
simultaneous motion of defects are displayed.}
\label{evl}
\end{figure}

In order to estimate the pinning barrier, we compute the $IV$ characteristics
and the critical current $I_c$, and probe their behaviors with the system size $L$.
The voltage across the system is given by the ac Josephson relation\cite{Tinkham}
\begin{equation}
\langle V \rangle = \frac{\hbar}{2eL}\left\langle
\sum_{x}\frac{d(\phi_{x,2}-\phi_{x,1})}{dt}\right\rangle
\end{equation}
and the resulting characteristics are displayed in Fig.~\ref{iv} for system size
$L=8$, $16$, and $24$.
Systems larger than $L=24$ turn out to exhibit the same $IV$ characteristics as the case
$L=24$ and are thus not shown here. It is observed that non-zero voltage develops
as the driving current $I$ is increased beyond a certain value.
The size-dependence of the corresponding critical current $I_c$ is plotted in the inset
of Fig.~\ref{iv}, which demonstrates that $I_c$ first reduces with the system size $L$ and
saturates to a nearly constant value beyond $L=24$.
In the thermodynamic limit, $I_c$ is shown to approach the value $0.089$
(in units of the single-junction critical current $i_c$);
this is close to the value $I_c \approx 0.1$ at $f=1/2$, extracted roughly from
Fig.~1(a) of Ref.~\onlinecite{Stroud}.

In Fig.~\ref{evt}, we display the typical time dependence of the energy $E(t)$.
With the driving current $I = I_c(L) + 0.0001$ just above the
critical value, the energy is calculated through the use of Eq.~\eqref{energy}.
Note in Fig.~\ref{evt} that (a) and (b) correspond to sequential motion of defects
for the system size $L = 8$ (smaller than $40$) and $64$ (larger than $40$), respectively.
As the defect moves across one plaquette, $E(t)$ goes through a maximum
corresponding to the excited state discussed in Sec. III.
The lowest-energy state corresponds to configuration G
and the maximum one to M shown in Fig.~\ref{thconf}.
As pointed out, the defects can move simultaneously for appropriate
initial conditions. Such simultaneous motion is indeed observed in Fig.~\ref{evt}(c),
which reveals the doubling of both the amplitude and the period of $E(t)$ (i.e.,
the energy barrier and the period of the defect motion).
The two transient states seen in the inset of Fig.~\ref{evt}(c) indicate
that the system possessing two defects is not completely coherent in the
first stage of the dynamics.

The pinning energy barrier $E_B$, defined to be the difference between the
maximum energy $E(M)$ and the minimum one $E(G)$, is thus computed
as the system size is varied.
The size dependence of $E_B$ is then examined and shown in Fig.~\ref{evl}
for (a) sequential and (b) simultaneous motion of defects.
In the former case, the energy barrier is observed to approach the value
\begin{equation}
E_B = 0.1827
\end{equation}
in the thermodynamic limit.
This value is slightly below the one found analytically in the eight-plaquette system,
as expected.
In the case of simultaneous motion, Fig.~\ref{evl}(b) shows that the energy barrier becomes double
for the system size $L>40$.

\section{Summary}

We have studied the dynamics of domain wall defects created by
adding an extra vortex in a fully frustrated Josephson-junction ladder.
The defects are in general pinned by the energy barrier
generated by the underlying lattice structure and other vortices induced by
an external magnetic field or frustration.
Making use of the symmetry and topological constraints, we have computed the energy
barrier $E_B$ in systems of size $L \leq 8$.
The defects may be put to motion by applying currents larger than the critical current.
The corresponding motion in the system, driven by
uniform currents just above the critical value, has been investigated
by means of dynamical simulations performed directly on the equations of motion.
The resulting numerical estimation of $E_B=0.1827$ (in units of the Josephson coupling energy)
is fully consistent with the analytical
value obtained from resolution of the phases in the eight-plaquette ($L=8$) system.
In the dynamical study of the system, we have also observed that the
defects move sequentially in small systems ($L< 40$).  On the other hand,
in larger systems, the domain walls may also display coherent motion,
namely, they can move simultaneously as well as sequentially,
depending on the initial configurations.
Such difference has been attributed to the distance-dependent interaction between
defects and the underlying lattice geometry.

\section*{Acknowledgments}

One of us (M.Y.C.) acknowledges the visitor grant from the CNRS, France and
thanks the Laboratoire de Physique Th\'eorique, Strasbourg, for its kind
hospitality during his stay.
H.K. thanks K. Lee for help in obtaining numerical solutions.
This work was supported in part by the Korea Science and Engineering Foundation
through National Core Research Center for Systems Bio-dynamics
and by the Ministry of Education of Korea through the BK21 program.


\begin{thebibliography}{99}

\bibitem{FXY}
S. Teitel and C. Jayaprakash, Phys. Rev. B {\bf 27}, R598 (1983);
M.Y. Choi and S. Doniach, {\it ibid}. {\bf 31}, 4516 (1985);
W.-Y. Shih and D. Stroud, {\it ibid}. {\bf 32}, 158 (1985).

\bibitem{review}
For a review, see, e.g.,
B.J. Kim, G.S. Jeon, and M.Y. Choi, in \textit{High-Temperature Superconductors Vol. 39:
Studies of Josephson Junction Arrays}, edited by A.V. Narlikar (Nova, New York, 2001), pp. 245-267.

\bibitem{dyn}
For some recent works on dynamics, see
G.S. Jeon, J.S. Lim, H.J. Kim, and M.Y. Choi,
Phys. Rev. B {\bf 66}, 024511 (2002);
G.S. Jeon, S.J. Lee, and M.Y. Choi,
{\it ibid}. {\bf 67}, 014501 (2003);
J.S. Lim, M.Y. Choi, J. Choi, and B.J. Kim,
{\it ibid}. {\bf 69}, 220504(R) (2004).
%

\bibitem{defect}
J.S. Lim, M.Y. Choi, B.J. Kim, and J. Choi,
Phys. Rev. B {\bf 71}, 100505(R) (2005).

\bibitem{exp}
M.S. Rzchowski, S.P. Benz, M. Tinkham, and C.J. Lobb, Phys. Rev. B {\bf 42}, 2041 (1990);
S.P. Benz, M.S. Rzchowski, M. Tinkham, and C.J. Lobb, {\it ibid}. {\bf 42}, 6165 (1990).
%


\bibitem{Mazo}
J.J. Mazo, F. Falo, and L.M. Flor\'ia, Phys. Rev. B {\bf 52}, 10433 (1995);
C. Denniston and C. Tang, Phys. Rev. Lett. {\bf 75}, 3930 (1995).

\bibitem{Stroud}
I.-J. Hwang, S. Ryu, and D. Stroud, Phys. Rev. B {\bf 53}, R506 (1996);
M. Barahona, S.H. Strogatz, and T.P. Orlando, {\it ibid}. {\bf 57}, 1181 (1998).

\bibitem{Kim}
S. Kim, Phys. Lett. A {\bf 229}, 190 (1997);
J. Yi, S.-I. Lee, and S. Kim, Phys. Rev. B {\bf 63}, 132501 (2001).

\bibitem{quantum}
M. Kardar, Phys. Rev. B {\bf 33}, 3125 (1986).
M. Lee, M.-S. Choi, and M.Y. Choi,
{\it ibid}. {\bf 68}, 144506 (2003).

\bibitem{SR}
B.J. Kim, M.-S. Choi, P. Minnhagen, G.S. Jeon, H.J. Kim, and M.Y. Choi,
Phys. Rev. B {\bf 63}, 104506 (2001);
G.S. Jeon and M.Y. Choi,
{\it ibid}. {\bf 66}, 064514 (2002).

\bibitem{Tinkham}
See, e.g., M. Tinkham, {\it Introduction to Superconductivity}, 2nd ed. (McGraw-Hill, New York, 1995).

\end{thebibliography}
\end{document}